\documentclass [a4paper,fleqn, 12pt]{article}
\usepackage{graphicx}
\usepackage {subfigure,epsfig}

\usepackage {amsmath} \usepackage{amssymb} \usepackage{cite} \usepackage{amsthm}
\numberwithin{equation}{section} \numberwithin{table}{section} \mathindent=0pt
\theoremstyle{plain} \newtheorem{theorem}{Theorem}
\newtheorem{lemma}{Lemma}

\numberwithin{theorem}{section}
 \numberwithin{lemma}{section}

\begin{document}

\title{\textbf{Explicit form of the Yablonskii - Vorob'ev polynomials}}

\author{Maria V. Demina, Nikolai A. Kudryashov}

\date{Department of Applied Mathematics\\
Moscow Engineering and Physics Institute\\ (State University)\\
31 Kashirskoe Shosse, 115409, Moscow, \\ Russian Federation}
\maketitle

\begin{abstract}
Special polynomials associated with rational solutions of the second
Painlev\'{e} equation and other members of its hierarchy are
discussed. New approach, which allows one to construct each
polynomial is presented. The structure of the polynomials is
established. Formulas of their coefficients are found. Correlations
between the roots of every polynomial are obtained.
\end{abstract}



\section{Introduction}

It is well known that the second Painlev\'{e} equation $(P_2)$
\begin{equation}
\label{1.1}w_{zz}=2w^3+zw+\alpha
\end{equation}
has rational solutions only at integer values of the parameter
$\alpha$ $(\alpha=n\in \textbf{Z})$. These solutions can be written
in terms of the Yablonskii -- Vorob'ev polynomials $Q_n(z)$
\cite{Yablonskii01, Vorob'ev01}
\begin{equation}
\begin{gathered}
\label{1.2}\hfill w(z;n)=\frac{d}{dz}\left\{\ln
\left[\frac{Q_{n-1}(z)}{Q_n(z)}\right]\right\},\quad\,
n\geq1,\quad\,\,w(z;-n)=-w(z;n).
\end{gathered}
\end{equation}
The polynomials $Q_n(z)$ satisfy the differential -- difference
equation
\begin{equation}
\label{1.3}Q_{n+1}Q_{n-1}=zQ_n^2-4(Q_nQ_n^{''}-(Q_n^{'})^2),
\end{equation}
where $Q_0(z)=1$, $Q_1(z)=z$. It is not clear from the first sight
that this relation defines exactly polynomials however it is so.
Moreover $Q_n(z)$ are monic polynomials with integer coefficients.
These polynomials can be regarded as nonlinear analogues of
classical special polynomials. They possess a certain number of
interesting properties. In particular, for every integer positive
$n$ the polynomial $Q_n(z)$ has simple roots only and besides that,
two successive polynomials $Q_n(z)$ and $Q_{n+1}(z)$ do not have a
common root. Partial solutions of the Korteveg -- de Vries equation,
the modified Korteveg -- de Vries equation, the nonlinear
Schr\"{o}dinger equation can be expressed via the polynomials
$Q_n(z)$ \cite{Clarkson01}.

One of the most important problems concerning the Yablonskii -
Vorob'ev polynomials includes constructing explicit formulas of
their coefficients \cite{Clarkson02}. The attempt of solving this
problem can be found in recent work \cite{Kaneko01}, where the
coefficient of the lowest degree term was discussed.

In this work we present a new method, which allows one to find
formulas for the coefficients of every polynomial, to determine the
polynomial structure and to obtain correlations between its roots.
Our approach can be also applied for constructing other polynomials
related to nonlinear differential equations. In particular, we will
briefly review the case of some other equations of the $P_2$
hierarchy.

The outline of this paper is as follows. In section 2 the algorithm
of our method is presented and main theorems are proved.
Correlations between the roots of the Yablonskii - Vorob'ev
polynomials are established in section 3. Formulas of coefficients
are found in section 4. The polynomials associated with the second
and the third equations of the $P_2$ hierarchy are discussed in
sections 5 and 6, accordingly.

\section{Method applied}

Being of degree $n(n+1)/2$ the polynomial $Q_n(z)$ can be written as

\begin{equation}
\label{1.4}Q_n(z)=\sum_{k=0}^{n(n+1)/2}A_{n,k}z^{n(n+1)/2-k},\qquad
A_{n,0}=1.
\end{equation}

Let us show that it is possible to obtain each polynomial without
leaning on the recurrence formula \eqref{1.3}. For this aim we will
use a power expansion at infinity for solutions of the equation
\eqref{1.1}. This expansion found in \cite{Bruno02} is the following

\begin{equation}
\label{1.5}w(z;\alpha)=\frac{c_{\alpha,-1}}{z}+\sum_{l=1}^{\infty}c_{\alpha,-3l-1}z^{-3l-1},
\quad z\rightarrow \infty.
\end{equation}

Here all the coefficients $c_{\alpha,-3l-1}$ can be sequently found.
Taking into account four members \eqref{1.5} can be written as

\begin{equation}
\begin{gathered}
\label{1.5_1}w(z;\alpha)=-{\frac {\alpha}{z}}+{\frac {2\alpha\,
\left( \alpha-1 \right)
 \left( \alpha+1 \right) }{{z}^{4}}}-{\frac {4\alpha\, \left( \alpha
-1 \right)  \left( \alpha+1 \right)  \left( 3\,{\alpha}^{2}-10
 \right) }{{z}^{7}}}+\\
{\frac {8\alpha\, \left( \alpha-1 \right)
 \left( \alpha+1 \right)  \left( 12\,{\alpha}^{4}-117\,{\alpha}^{2}+
280 \right) }{{z}^{10}}}+O(\frac1{z^{13}}).
\end{gathered}
\end{equation}

For convenience of use let us present this series in the form

\begin{equation}
\label{1.5a}w(z;\alpha)=\sum_{m=0}^{\infty}c_{\alpha,-(m-1)}z^{-m-1},
\end{equation}

where $c_{\alpha,-(m-1)}=0$ unless $m$ is divisible by $3$. Suppose
$a_{n,k}$ $(1\leq k\leq n(n+1)/2)$ are the roots of the polynomial
$Q_n(z)$, then by $s_{n,k}$ we denote the symmetric functions of the
roots

\begin{equation}
\label{1.7}s_{n,m}\stackrel{def}{=}\sum_{k=1}^{n(n+1)/2}(a_{n,k})^m,\quad
m\geq 1.
\end{equation}

Our next step is to express $s_{n,m}$ through coefficients of the
series \eqref{1.5a}.

\begin{theorem}
\label{L:1} Let $c_{i,-m-1}$ be the coefficient in expansion
\eqref{1.5a} at integer $\alpha=i\in \textbf{N}$. Then for each
$m\geq1$ and $n\geq2$ the following relation holds

\begin{equation}
\label{1.8}s_{n,m}=-\sum_{i=2}^{n}c_{i,-(m+1)}.
\end{equation}
\end{theorem}

\begin{proof}
As far as $Q_n(z)$ is a monic polynomial with simple roots, then it
can be written in the form

\begin{equation}
\label{1.9}Q_n(z)=\prod_{k=1}^{n(n+1)/2}(z-a_{n,k}).
\end{equation}

This implies that

\begin{equation}
\label{1.10}\frac{Q_n^{'}(z)}{Q_n(z)}=\sum_{k=1}^{n(n+1)/2}\frac1{z-a_{n,k}}.
\end{equation}

Substituting \eqref{1.10} into the expression \eqref{1.2} yields

\begin{equation}
\label{1.11}w(z;n)=\frac{Q_{n-1}^{'}(z)}{Q_{n-1}(z)}-\frac{Q_n^{'}(z)}{Q_n(z)}=
\sum_{k=1}^{n(n-1)/2}\frac1{z-a_{{n-1},k}}-\sum_{k=1}^{n(n+1)/2}\frac1{z-a_{n,k}}.
\end{equation}

Expanding this function in a neighborhood of infinity we get

\begin{equation}
\begin{gathered}
\label{1.12}w(z;n)=\sum_{k=1}^{n(n-1)/2}\frac1{z(1-\frac{a_{{n-1},k}}{z})}-
\sum_{k=1}^{n(n+1)/2}\frac1{z(1-\frac{a_{n,k}}{z})}=
\frac{\delta_{0,a_{n-1,1}}}{z}-\frac{\delta_{0,a_{n,1}}}{z}+\\
\\
\sum_{m=0}^{\infty}
\left[\sum_{k=1+\delta_{0,a_{n-1,1}}}^{n(n-1)/2}(a_{n-1,k})^m-\sum_{k=1+
\delta_{0,a_{n,1}}}^{n(n+1)/2}(a_{n,k})^m\right]z^{-(m+1)},\,\,
|z|>\max\{\tilde{a}_{n-1},\tilde{a}_{n} \},\\
\\
\tilde{a}_{n-1}=\max\limits_{1\leq k \leq
n(n-1)/2}\{|a_{n-1,k}|\},\,\tilde{a}_{n}=\max\limits_{1\leq k \leq
n(n+1)/2}\{|a_{n,k}|\},
\end{gathered}
\end{equation}

where the first or the second term is present only if the polynomial
$Q_{n-1}(z)$ or $Q_n(z)$ has a zero root, which without loss of
generality is the first in the set of roots. In our designations the
previous expression can be rewritten as

\begin{equation}
\begin{gathered}
\label{1.13}w(z;n)=-\frac{n}{z}+\sum_{m=1}^{\infty}\left[s_{n-1,m}-
s_{n,m}\right]z^{-(m+1)},\,\,|z|>\max\{\tilde{a}_{n-1},\tilde{a}_{n}\}.
\end{gathered}
\end{equation}

The absence of a zero term in sum is essential only at $m=0$.
Comparing expansions \eqref{1.13} and \eqref{1.5a} we obtain the
equality

\begin{equation}
\label{1.14}s_{n,m}-s_{n-1,m}=-c_{n,-(m+1)}.
\end{equation}

Decreasing the first index by one in \eqref{1.14} and adding the
result to the original one yields

\begin{equation}
\label{1.15}s_{n,m}-s_{n-2,m}=-(c_{n,-(m+1)}+c_{n-1,-(m+1)}).
\end{equation}

Note that $c_{1,-(m+1)}=0,\,m\geq1$ and $a_{1,1}=0$. Then proceeding
in such a way we get the required relation \eqref{1.8}.
\end{proof}

\textit{Remark 1.} It was proved many times that $P_2$ has a unique
rational solution whenever $\alpha$ is an integer. All these
solutions possess convergent series at infinity. Note that every
rational solution $w(z;n)$ has the asymptotic behavior
\begin{equation*}
\label{1.15}w(z;n)\backsim -\frac{n}{z},\quad z\rightarrow \infty,
\end{equation*}
i.e. the point $z=\infty$ is a simple root. This fact can be easily
seen from \eqref{1.13}. Thus the formal series \eqref{1.5} at
$\alpha=n$ coincides with the expansion \eqref{1.12} and is also
convergent.

Theorem \eqref{L:1} enables us to prove the following theorem.
\begin{theorem}
\label{T:1.1.} All the coefficients $A_{n,m}$ of the Yablonskii --
Vorob'ev polynomial $Q_n(z)$ can be obtained with a help of
$n(n+1)/2+1$ first coefficients of the expansion \eqref{1.5a} for
the solutions of $P_2$.
\end{theorem}

\begin{proof} For every polynomial there exists a connection between
its coefficients and the symmetric functions of its roots $s_{n,m}$.
This connection is the following
\begin{equation}
\label{1.16}mA_{n,m}+s_{n,1}A_{n,m-1}+\ldots +s_{n,m}A_{n,0}=0,\quad
1\leq m\leq n(n+1)/2.
\end{equation}
Taking into account that $A_{n,0}=1$ we get
\begin{equation}
\label{1.17}A_{n,m}=-\frac{s_{n,m}+s_{n,m-1}A_{n,1}+\ldots
+s_{n,1}A_{n,m-1}}{m},\quad 1\leq m\leq n(n+1)/2.
\end{equation}
The function $s_{n,m}$ can be derived using the expression
\eqref{1.8}. Hence recalling the fact that \eqref{1.5a} is exactly
\eqref{1.5} we obtain
\begin{equation}
\begin{gathered}
\label{1.18}s_{n,m}=0,\quad m\in \textbf{N}\,/\,\{3l,\quad l\in
\textbf{N}\},\\
s_{n,3l}=-\sum_{i=2}^{n}c_{i,-(3l+1)},\quad l\in \textbf{N}.
\end{gathered}
\end{equation}
Substituting this into \eqref{1.17} yields
\begin{equation}
\begin{gathered}
\label{1.19}A_{n,m}=0,\quad m\in\{1,2,\ldots
,n(n+1)/2\}\,/\,\{3l,\quad l\in \textbf{N}\};\\
\\
A_{n,3l}=-\frac{s_{n,3l}+s_{n,3l-3}A_{n,3}+\ldots
+s_{n,3}A_{n,3l-3}}{3l},\quad l\in \textbf{N},\,3l\leq n(n+1)/2.
\end{gathered}
\end{equation}
Thus we see that the coefficients $A_{n,k}$ of the Yablonskii --
Vorob'ev polynomial $Q_n(z)$ are uniquely defined by coefficients
$c_{n,-(3l+1)}$ of the expansion \eqref{1.5}. This completes the
proof.
\end{proof}
\textit{Remark 2.} Expression \eqref{1.19} defines the structure of
polynomial $Q_n(z)$. Namely if $n(n+1)/2$ is divisible by $3$, i.e.
$n\equiv0\,mod\, 3$ or $n\equiv2\,mod\, 3$, then $Q_n(z)$  is a
polynomial in $z^3$. Otherwise if $n(n+1)/2$ is not divisible by
$3$, i.e. $n\equiv1\,mod\, 3$, then the absolute term of $Q_n(z)$ is
equal to zero and $Q_n(z)/z$ is a polynomial in $z^3$ (as in this
case $n(n+1)/2-1$ is divisible by 3).

\section{Symmetric functions of the roots}

In this section we are discussing properties of the functions
$s_{n,m}$. It is important to mention that they may be regarded as
relations between the roots $a_{n,k}$ of the Yablonskii - Vorob'ev
polynomials. In order to establish our main results we need a lemma.
\begin{lemma}
\label{L:2.1} The coefficient $c_{\alpha,-3l-1}$ in the expansion
\eqref{1.5} is a polynomial in $\alpha$ of degree $2l+1$.
\end{lemma}
\begin{proof}
The proof is by induction on $l$. For $l=0$, there is nothing to
prove as $c_{\alpha,-1}=-\alpha$. Other coefficients can be obtained
from the recursion relation
\begin{equation}\begin{gathered}
\label{2.1}c_{\alpha,-3(l+1)-1}=(3l+2)(3l+1)c_{\alpha,-3l-1}-2\sum_{m=0}^{l}\sum_{n=0}^{m}
c_{\alpha,-3n-1}\\
c_{\alpha,-3(m-n)-1}c_{\alpha,-3(l-m)-1},\quad l\geq0.
\end{gathered}\end{equation}
Suppose that $c_{\alpha,-3m-1}$ is a polynomial in $\alpha$ of
degree $2m+1$ $(0<m\leq l)$. Then from \eqref{2.1} we see that
$c_{\alpha,-3(l+1)-1}$ is also a polynomial in $\alpha$ and $deg(
c_{\alpha,-3(l+1)-1})=2n+1+2(m-n)+1+2(l-m)+1=2l+3=2(l+1)+1$. Q.E.D.
\end{proof}

\begin{theorem}
\label{T:2.1} The following statements are true:

1. at given $n\geq2$ the functions $s_{n,m}$ $(m>n(n+1)/2)$ do not
contain any new information about the roots of $Q_n(z)$;

2. $s_{n,3l}$ is a polynomial in $n$ of degree $2(l+1)$.
\end{theorem}
\begin{proof}
The first statement of the theorem immediately follows from the
correlation
\begin{equation}
\label{2.2}s_{n,m}+s_{n,m-1}A_{n,1}+\ldots
+s_{n,m-n(n+1)/2}A_{n,n(n+1)/2}=0,\quad m>n(n+1)/2
\end{equation}
and the expression \eqref{1.17}. Now let us prove the second
statement. From \eqref{1.8} and Lemma \eqref{L:2.1} we see that in
order to find $s_{n,3l}$ one should calculate finite amount of sums
$\sum\limits_{i=1}^{n}i^m$, $m\in \textbf{N}$, $\max m=2l+1$. Such
sum is computable. And the result is a polynomial in $n$ of degree
$m+1$. This completes the proof.
\end{proof}
Finally let us find several functions $s_{n,3l}$. They are

\begin{equation}\begin{gathered}
\label{k1}s_{{{n,3}}}=-\frac12\,n \left( {n}^{2}-1 \right) \left(
n+2 \right),
\end{gathered}\end{equation}

\begin{equation}\begin{gathered}
\label{k2}s_{{{n,6}}}=2\,n \left( {n}^{2}-1 \right)  \left( n+2
\right)
 \left( {n}^{2}+n-5 \right),
\end{gathered}\end{equation}

\begin{equation}\begin{gathered}
\label{k3}s_{{{n,9}}}=-4\,n \left( {n}^{2}-1 \right)  \left( n+2
\right)
 \left( {n}^{2}+n-7 \right)  \left( 3\,{n}^{2}+3\,n-20 \right),
\end{gathered}\end{equation}

\begin{equation}\begin{gathered}
\label{k4}s_{{{n,12}}}=8\,n \left( {n}^{2}-1 \right)  \left( n+2
\right)
 [11\,{n}^{6}+33\,{n}^{5}-259\,{n}^{4}-573\,{n}^{3}+\\
 \\
 +2348\,{n}^{2
}+2640\,n-7700 ],
\end{gathered}\end{equation}

\begin{equation}\begin{gathered}
\label{k5}s_{{{n,15}}}=-8\,n \left( {n}^{2}-1 \right)  \left( n+2
\right)
 [91\,{n}^{8}+364\,{n}^{7}-3468\,{n}^{6}-11678\,{n}^{5}+\\
\\
+57138\,{ n}^{4}+134164\,{n}^{3}-454161\,{n}^{2}-523250\,n+1401400],
\end{gathered}\end{equation}

\begin{equation}\begin{gathered}
\label{k6}s_{{{n,18}}}=32\,n \left( {n}^{2}-1 \right)  \left( n+2
\right)
 [204\,{n}^{10}+1020\,{n}^{9}-11584\,{n}^{8}-\\
 \\
 -52456\,{n}^{7}+303649\,{n}^{6}+1098827\,{n}^{5}-4328687\,{n}^{4}-\\
\\
-10551991\,{n}^{3}+ 32064418\,{n}^{2}+37532600\,n-95295200],
\end{gathered}\end{equation}

\begin{equation}\begin{gathered}
\label{k7}s_{{{n,21}}}=-64\,n \left( {n}^{2}-1 \right)  \left( n+2
\right)
 [969\,{n}^{12}+5814\,{n}^{11}-77478\,{n}^{10}-\\
 \\
-440685\,{n}^{9}+
2986374\,{n}^{8}+14653560\,{n}^{7}-66988510\,{n}^{6}-\\
\\
-254167821\,{n}^{5
}+882020165\,{n}^{4}+2205662532\,{n}^{3}-\\
\\
-6249767920\,{n}^{2}- 7397539800\,n+18106088000],
\end{gathered}\end{equation}

\begin{equation}\begin{gathered}
\label{k8}s_{{{n,24}}}=128\,n \left( n-1 \right)  \left( n+2 \right)
\left( n+1 \right) [4807\,{n}^{14}+33649\,{n}^{13}-\\
\\
-519091\,{n}^{12}-3551983\,{n}^{11}+27883155\,{n}^{10}+172777587\,{n}^{9}-\\
\\
-911403269\,{n} ^{8}-4722725213\,{n}^{7}+18734279962\,{n}^{6}+\\
\\
+73498559352\,{n}^{5}-234405400524\,{n}^{4}-597184066192\,{n}^{3}+\\
\\
+1610723930960\,{n}^{2}+1922407748800\,n-4580840264000],
\end{gathered}\end{equation}

\begin{equation}\begin{gathered}
\label{k9}s_{{{n,27}}}=-128\,n \left( n^2-1 \right)  \left( n+2
\right) [49335\,{n}^{16}+394680\,{n}^{15}-\\
\\
-6966616\, {n}^{14}-55673212\,{n}^{13}+501236020\,{n}^{12}+3749125816\,{n}^{11}-\\
\\
-22662344352\,{n}^{10}-149052619326\,{n}^{9}+674231177321\,{n}^{8}+\\
\\
+3634840116452\,{n}^{7}-13091078484596\,{n}^{6}-52664915417010\,{n}^{5}+\\
\\
+158038744882088\,{n}^{4}+408412972732600\,{n}^{3}-\\
\\
-1061767782349200\,{ n}^{2}-1275405591460000\,n+\\
\\
+2977546171600000].
\end{gathered}\end{equation}

\section{Coefficients of the Yablonskii - Vorob'ev polynomials}

Before coming to the direct computation of the coefficients it is
important to mention that $A_{n,3l}$ is a polynomial in $n$ of
degree $4l$. This fact can be proved by induction. Using the results
of the previous section and the expression \eqref{1.19} we obtain

\begin{equation}\begin{gathered}
\label{k11}A_{{{n,3}}}=\frac{n}{6}\,\left( {n}^{2}-1 \right) \left(
n+2 \right),
\end{gathered}\end{equation}

\begin{equation}\begin{gathered}
\label{k12}A_{{{n,6}}}={\frac {n}{72}}\,\left( {n}^{2}-1 \right)
\left( {n} ^{2}-4 \right)  \left( n-4 \right)  \left( n+5 \right)
\left( n+3
 \right),
\end{gathered}\end{equation}

\begin{equation}\begin{gathered}
\label{k13}A_{{{n,9}}}={\frac {n}{1296}}\,\left( {n}^{2}-1 \right)
\left( { n}^{2}-4 \right)  \left( {n}^{2}-9 \right)  \left( n+4
\right)
 \\
 \\
 \left( {n}^{4}+2\,{n}^{3}-57\,{n}^{2}-58\,n+1120 \right),
\end{gathered}\end{equation}

\begin{equation}\begin{gathered}
\label{k14}A_{{{n,12}}}={\frac {n}{31104}}\,\left( {n}^{2}-1 \right)
 \left( {n}^{2}-4 \right)  \left( {n}^{2}-9 \right)  \left( {n}^{2}-16
 \right)  \left( n+5 \right)  \\
 \\
 \left( {n}^{6}+3\,{n}^{5}-109\,{n}^{4}-
223\,{n}^{3}+5148\,{n}^{2}+5260\,n-110880 \right),
\end{gathered}\end{equation}

\begin{equation}\begin{gathered}
\label{k15}A_{{{n,15}}}={\frac {n}{933120}}\,\left( {n}^{2}-1
\right)
 \left( {n}^{2}-4 \right)  \left( {n}^{2}-9 \right)  \left( {n}^{2}-16
 \right)  \left( n+5 \right) [{n}^{10}+\\
 \\
 +5\,{n}^{9} -200\,{n}^{8}-
830\,{n}^{7}+18917\,{n}^{6}+59677\,{n}^{5}-1072550\,{n}^{4}-\\
\\
-2245540\,{ n}^{3}+35648392\,{n}^{2}+36781248\,n-484323840],
\end{gathered}\end{equation}

\begin{equation}\begin{gathered}
\label{k16}A_{{{n,18}}}={\frac {n}{33592320}}\,\left( {n}^{2}-1
\right)
 \left( {n}^{2}-4 \right)  \left( {n}^{2}-9 \right)  \left( {n}^{2}-16
 \right)  \left( {n}^{2}-25 \right) \\
 \\
  \left( n+6 \right) [{n}^{
12}+6\,{n}^{11}-287\,{n}^{10}-1490\,{n}^{9}+40087\,{n}^{8}+169354\,{n}
^{7}-\\
\\
-3558197\,{n}^{6}-11273654\,{n}^{5}+213009052\,{n}^{4}+445008120\,
{n}^{3}-\\
\\
-7958672096\,{n}^{2}-8183083776\,n+131736084480],
\end{gathered}\end{equation}

\begin{equation}\begin{gathered}
\label{k17}A_{{{n,21}}}={\frac {n}{1410877440}}\, \left( {n}^{2}-1
\right)
 \left( {n}^{2}-4 \right)  \left( {n}^{2}-9 \right)  \left( {n}^{2}-16
 \right)  \left( {n}^{2}-25 \right) \\
 \\
 \left( n+6 \right)  [{n}^{
16}+8\,{n}^{15}-420\,{n}^{14}-3080\,{n}^{13}+87206\,{n}^{12}+563640\,{
n}^{11}-\\
\\
-11872404\,{n}^{10}-64603080\,{n}^{9}+1168725105\,{n}^{8}+5068846496\,{n}^{7}-\\
\\
-84738361232\,{n}^{6}-272233713360\,{n}^{5}+4305162496688\,{n}^{4}+\\
\\
+9070093877056\,{n}^{3}-132946555052544\,{n}^{2}-137527838945280\,n+\\
\\
+1802149635686400],
\end{gathered}\end{equation}

\begin{equation}\begin{gathered}
\label{k18}A_{n,24}={\frac {n}{67722117120}}\,({n}^{2}-1)
({n}^{2}-4) ({n}^{2}-9)({n}^{2}-16)({n}^{2}-25)\\
 \\
 ({n}^{2}-36)\,( {n}^{2}-49)\,( n+8)\, [{n}^{16}+8\,{n}^{15}-492\,
{n}^{14}-3584\,{n}^{13}+\\
\\
+121478\,{n}^{12}+775824\,{n}^{11}-20020068\,{n
}^{10}-107298432\,{n}^{9}+\\
\\
+2443900401\,{n}^{8}+10428074408\,{n}^{7}-
227804919608\,{n}^{6}-\\
\\
-720372362304\,{n}^{5}+15749707896080\,{n}^{4}+32712421425280\,{n}^{3}-\\
\\
-723511683854592\,{n}^{2}-739989727488000\,n+ 16283709208166400],
\end{gathered}\end{equation}

\begin{equation}\begin{gathered}
\label{k19}A_{{{n,27}}}={\frac {n}{3656994324480}}\,\left( {n}^{2}-1
 \right)  \left( {n}^{2}-4 \right)  \left( {n}^{2}-9 \right)  \left( {
n}^{2}-16 \right)  \\
\\
\left( {n}^{2}-25 \right) \left( {n}^{2}-36
 \right)  \left( n+7 \right)\,[{n}^{22}+11\,{n}^{21}-715\,{n}^{
20}-7535\,{n}^{19}+\\
\\
+255960\,{n}^{18} +2520582\,{n}^{17}-61352518\,{n}^{
16}-548941070\,{n}^{15}+\\
\\
+11058177409\,{n}^{14}+87218618983\,{n}^{13}-1583236002911\,{n}^{12}-\\
\\
10658996892035\,{n}^{11}+183545605960118\,{n}^ {10}+1017581723269944\,{n}^{9}-\\
\\
-17057801205684864\,{n}^{8}-74457629190856880\,{n}^{7}+\\
\\
+1221271897326432992\,{n}^{6}+3928812356965402880\,{n}^{5}-\\
\\
-62376250932962964992\,{n}^{4}-131389485437862420480\,{n}^{3}+\\
\\
+1976136057743843819520\,{n}^{2}+2042498326581057945600\,n-\\
\\
-28450896728508334080000].
\end{gathered}\end{equation}

Now let us write out the general form of the Yablonskii - Vorob'ev
polynomial $Q_n(z)$. It is the following

\begin{equation}\begin{gathered}
\label{k100}{Q_n}( z) ={z}^{{n( n+1)}/{2}}+A_{n,3}\,{z}^{n( n+1)/{2}
-3}+A_{n,6}\, {z}^{n(n+1)/{2} -6}+\\
\\
+A_{n,9}\, {z}^{n( n+1)/{2}-9}+A_{n,12}\,{z}^{n(
n+1)/{2}-12}+A_{n,15}\,{z}^{n( n+1)/{2}-15}+\\
\\
+ +A_{n,18}\, {z}^{n( n+1)/{2}-18}+A_{n,21}\,{z}^{n(
n+1)/{2}-21}+A_{n,24}\, {z}^{n( n+1)/{2}-24}+\\
\\
+A_{n,27}\,{z}^{n( n+1)/{2}-27}+...
\end{gathered}\end{equation}
where the first nine coefficients $A_{n,3k}$, $(k=1,...,9)$ we have
just found (see \eqref{k11} - \eqref{k19}). Substituting
$n=0,\,1,\,2,\,3,\,4,\,5,\,6,\,7$ into expression \eqref{k100} one
can obtain the Yablonskii - Vorob'ev polynomials $Q_0(z)$, $Q_1(z)$,
$Q_2(z)$, $Q_3(z)$, $Q_4(z)$, $Q_5(z)$, $Q_6(z)$, and $Q_7(z)$.
Rational solutions of the second Painlev\'{e} equation \eqref{1.1}
can be found using the correlation \eqref{1.2}.

\section{Special polynomials associated with the second equation of the $P_2$ hierarchy}

In this section our interest is in the polynomials $Q_n^{[2]}(z)$
associated with the forth-order analogue to the second Painlev\'{e}
equation
\begin{equation}
\label{kk0}w_{zzzz}-10\,w^2\,w_{zz}-10\,w\,w_z^2+6\,w^5-z\,w-\alpha=0.
\end{equation}
Power expansion of its solutions in a neighborhood of infinity was
found in \cite{Demina01} and is the following
\begin{equation}
\label{kk0a}w(z;\alpha)=\frac{c_{\alpha,-1}}{z}+\sum_{l=1}^{\infty}c_{\alpha,-5l-1}z^{-5l-1},
\quad z\rightarrow \infty.
\end{equation}
Again for convenience of use let us rewrite this series in the form
\begin{equation}
\label{kk0b}w(z;\alpha)=\sum_{m=0}^{\infty}c_{\alpha,-(m-1)}z^{-m-1},
\end{equation}
where $c_{\alpha,-(m-1)}=0$ unless $m$ is divisible by $5$. The
polynomial $Q_n^{[2]}(z)$ is a monic polynomial of degree $n(n+1)/2$
\begin{equation}
\label{1.4}Q_n^{[2]}(z)=\sum_{k=0}^{n(n+1)/2}A_{n,k}z^{n(n+1)/2-k},\qquad
A_{n,0}=1.
\end{equation}
Symmetric functions of its roots can be defined as it was done in
the case of the Yablonskii-Vorob'ev polynomials
\begin{equation}
\label{kk0c}s_{n,m}\stackrel{def}{=}\sum_{k=1}^{n(n+1)/2}(a_{n,k})^m,\quad
m\geq 1,
\end{equation}
where $a_{n,k}$ $(1\leq k\leq n(n+1)/2)$ are the roots of
$Q_n^{[2]}(z)$. The following theorems are true:

\begin{theorem}
\label{T:5.1} Let $c_{i,-m-1}$ be the coefficient in expansion
\eqref{kk0b} at integer $\alpha=i\in \textbf{N}$. Then for each
$m\geq1$ and $n\geq2$ the following relation holds
\begin{equation}
\label{kk0d}s_{n,m}=-\sum_{i=2}^{n}c_{i,-(m+1)}.
\end{equation}
\end{theorem}

\begin{theorem}
\label{T:5.2} All the coefficients $A_{n,m}$ of the polynomial
$Q_n^{[2]}(z)$ can be obtained with a help of $n(n+1)/2+1$ first
coefficients of the expansion \eqref{kk0b} for the solutions of
\eqref{kk0}.
\end{theorem}

These theorems can be proved in the same way as in section 2. Using
the expression \eqref{kk0d} we get

\begin{equation}\begin{gathered}
\label{kk1}s_{{{n,5}}}=n \left( {n}^{2}-1 \right) \left( {n}^{2}-4,
\right)
 \left( n+3 \right)
\end{gathered}\end{equation}

\begin{equation}\begin{gathered}
\label{kk2}s_{{{n,10}}}=6\,n \left({n}^{2}-1 \right)\left( {n}^{2}-4
\right)\left( n+3
\right)[3\,{n}^{4}+6\,{n}^{3}-73\,{n}^{2}-76\,n+504],
\end{gathered}\end{equation}

\begin{equation}\begin{gathered}
\label{kk3}s_{{{n,15}}}=36\,n \,\left( {n}^{2}-1 \right) \left(
{n}^{2}-4 \right)  \left( n+3 \right)\,\, [
15\,{n}^{8}+60\,{n}^{7}-1010\,{n}
^{6}-\\
\\
-3240\,{n}^{5}+28759\,{n}^{4}+62988\,{n}^{3}-388124\,{n}^{2}-
420168\,n+2018016 ],
\end{gathered}\end{equation}

\begin{equation}\begin{gathered}
\label{kk4}s_{{{n,20}}}=72\,n \left( {n}^{2}-1 \right) \left(
{n}^{2}-4
 \right)  \left( n+3 \right)\, [285\,{n}^{12}+1710\,{n}^{11}-\\
 \\
 -37965\,{n}^{10}-205500\,{n}^{9}+2387695\,{n}^{8}+10802590\,{n}^{7}-
85963355\,{n}^{6}-\\
\\
296581040\,{n}^{5} +1799452252\,{n}^{4}+4106229664\,{
n}^{3}-\\
\\
-20241225792\,{n}^{2} -22345634304\,n+93861960192],
\end{gathered}\end{equation}

\begin{equation}\begin{gathered}
\label{kk5}s_{{{n,25}}}=864\,n \left( {n}^{2}-1 \right) \left(
{n}^{2}-4
 \right)  \left( n+3 \right) [1035\,{n}^{16}+8280\,{n}^{15}-\\
 \\
 -233460\,{n}^{14}-1779120\,{n}^{13}+26279210\,{n}^{12}+181180560\,{n}^{
11}-\\
\\
-1828510100\,{n}^{10}-10846761360\,{n}^{9}+82823297235\,{n}^{8}+\\
\\
+398441209080\,{n}^{7}-2435998476560\,{n}^{6}-8750167253280\,{n}^{5}+\\
\\
+44588973389072\,{n}^{4}+104249136461184\,{n}^{3}-457807824496512\,{n}^
{2}-\\
\\
-511460571815424\,n+1994754378000384],
\end{gathered}\end{equation}

\begin{equation}\begin{gathered}
\label{kk6}s_{{{n,30}}}=864\,n \left( {n}^{2}-1 \right) \left(
{n}^{2}-4
 \right)  \left( n+3 \right) [49329\,{n}^{20}+493290\,{n}^{19}-\\
 \\-
17146575\,{n}^{18}-168377940\,{n}^{17}+3079121634\,{n}^{16}+
28513272780\,{n}^{15}-\\
\\
-357124460950\,{n}^{14}-3012530148880\,{n}^{13}+
28532122080349\,{n}^{12}+\\
\\
+211703836490170\,{n}^{11}-1597880669280075\,{
n}^{10}-\\
\\
-10005358208913420\,{n}^{9}+62322893943391984\,{n}^{8}+\\
\\
+ 311781921068301760\,{n}^{7}-1647676490226842800\,{n}^{6}-\\
\\
-6078753435828084160\,{n}^{5}+27931267677418875264\,{n}^{4}+\\
\\
+66378780435551109120\,{n}^{3}-271399919715872962560\,{n}^{2}-\\
\\
+305657587619581317120\,n+1137057869565290889216]
\end{gathered}\end{equation}

First several coefficients of the polynomials $Q_n^{[2]}(z)$ are the
following
\begin{equation}\begin{gathered}
\label{kk11}A_{{{n,5}}}=-\frac{n}{5}\left( {n}^{2}-1 \right) \left(
{n}^{2}-4
 \right)  \left( n+3 \right),
\end{gathered}\end{equation}

\begin{equation}\begin{gathered}
\label{kk12}A_{{{n,10}}}={\frac {n}{50}}\,\left( {n}^{2}-1 \right)
\left( {n }^{2}-4 \right)  \left( {n}^{2}-9 \right)  \left( n+4
\right) \\
\\
\left( {n}^{4}+2\,{n}^{3}-85\,{n}^{2}-86\,n+1260 \right),
\end{gathered}\end{equation}

\begin{equation}\begin{gathered}
\label{kk13}A_{{{n,15}}}=-{\frac {n}{750}}\,\left( {n}^{2}-1 \right)
\left( {n}^{2}-4 \right)  \left( {n}^{2}-9 \right)  \left(
{n}^{2}-16
\right)  \left( n+5 \right)\\
\\
[{n}^{8}+4\,{n}^{7}-248\,{n}^{6}-
758\,{n}^{5}+26959\,{n}^{4}+55186\,{n}^{3}-1107792\,{n}^{2}-\\
\\
-1135512\,n +15135120],
\end{gathered}\end{equation}

\begin{equation}\begin{gathered}
\label{kk14}A_{{{n,20}}}={\frac {1}{15000}}\,n \left( {n}^{2}-1
\right)
 \left( {n}^{2}-4 \right)  \left( {n}^{2}-9 \right)  \left( {n}^{2}-16
 \right)  \left( {n}^{2}-25 \right)  \\
 \\
 \left( n-7 \right) \left( n+8
 \right)  \left( n+6 \right)  [{n}^{10}+5\,{n}^{9}-436\,{n}^{8}-
1774\,{n}^{7}+94877\,{n}^{6}+\\
\\
+290861\,{n}^{5}-11996834\,{n}^{4}-
24480516\,{n}^{3}+661271112\,{n}^{2}+\\
\\
+673560144\,n-12570798240],
\end{gathered}\end{equation}

\begin{equation}\begin{gathered}
\label{kk15}A_{{{n,25}}}=-{\frac {n}{375000}}\,\left( {n}^{2}-1
\right)
 \left( {n}^{2}-4 \right)  \left( {n}^{2}-9 \right)  \left( {n}^{2}-16
 \right)  \left( {n}^{2}-25 \right) \\
 \\
 \left( {n}^{2}-36 \right)
 \left( n+7 \right) [{n}^{16}+8\,{n}^{15}-800\,{n}^{14}-5740\,{
n}^{13}+329098\,{n}^{12}+\\
\\
+2049572\,{n}^{11}-89974936\,{n}^{10}-
468800180\,{n}^{9}+16974375821\,{n}^{8}+\\
\\
+70733085892\,{n}^{7}-
2040797018848\,{n}^{6}-6371948611280\,{n}^{5}+\\
\\
+144781627711680\,{n}^{4}
+300266640707328\,{n}^{3}-\\
\\
-5485849799351616\,{n}^{2}-5637057042355200\, n+\\
\\
+85489473342873600],
\end{gathered}\end{equation}

\begin{equation}\begin{gathered}
\label{kk15a}A_{n,30}={\frac {n}{11250000}}\,\left( {n}^{2}-1
\right)  \left( {n}^{2}-4
 \right)  \left( {n}^{2}-9 \right)  \left( {n}^{2}-16 \right)  \left(
{n}^{2}-25 \right) \\
\\
 \left( {n}^{2}-36 \right)  \left( {n}^{2}-49
 \right)  \left( n+8 \right) [{n}^{20}+10\,{n}^{19}-1185\,{n}^{
18}-10950\,{n}^{17}+\\
\\
+732102\,{n}^{16}+6106308\,{n}^{15}-311393810\,{n}^
{14}-2287492220\,{n}^{13}+\\
\\
+99293702253\,{n}^{12}+625780701186\,{n}^{11}
-23733970000125\,{n}^{10}-\\
\\
-124461428270910\,{n}^{9}+4030862821261084\,{
n}^{8}+\\
\\
+16877198228842864\,{n}^{7}-456671374284826080\,{n}^{6}-\\
\\-1429614037781766720\,{n}^{5}+32300111441025610560\,{n}^{4}+\\
\\
+ 67002855424545877632\,{n}^{3}-1282130994934101484800\,{n}^{2}-\\
\\
-1315873542386590387200\,n+21754933728927759360000]
\end{gathered}\end{equation}

Rational solutions of \eqref{kk0} can be expressed via the
logarithmic derivative of the polynomials $Q_n^{[2]}(z)$
\begin{equation}
\begin{gathered}
\label{kk18}\hfill w(z;n)=\frac{d}{dz}\left\{\ln
\left[\frac{Q_{n-1}^{(2)}(z)}{Q_n^{(2)}(z)}\right]\right\},\quad\,
n\geq1,\quad\,\,w(z;-n)=-w(z;n).
\end{gathered}
\end{equation}

\section{Special polynomials associated with the third equation of the $P_2$ hierarchy}

In this section we will deal with the polynomials $Q_n^{[3]}(z)$
associated with the sixth-order analogue to the second Painlev\'{e}
equation
\begin{equation}\begin{gathered}
\label{6.1}w_{zzzzzz}-14\,{w}^{2}w_{{{zzzz}}}-56\,ww_{{z}}w_{{{
zzz}}}-42\,w{w_{{{zz}}}}^{2}-70\,{w_{{z}}}^{2}w_{{{zz}}}+\\
\\
+70\,{
w}^{4}w_{{{zz}}}+140\,{w}^{3}{w_{{z}}}^{2}-20\,{w}^{7}-z\,w-\alpha=0.
\end{gathered}
\end{equation}
Power expansion of its solutions in a neighborhood of infinity can
be presented in the form
\begin{equation}
\label{6.2}w(z;\alpha)=\frac{c_{\alpha,-1}}{z}+\sum_{l=1}^{\infty}c_{\alpha,-7l-1}z^{-7l-1},
\quad z\rightarrow \infty.
\end{equation}
Again for convenience of use let us rewrite this series as
\begin{equation}
\label{6.3}w(z;\alpha)=\sum_{m=0}^{\infty}c_{\alpha,-(m-1)}z^{-m-1},
\end{equation}
where $c_{\alpha,-(m-1)}=0$ unless $m$ is divisible by $7$. Since
the polynomial $Q_n^{[3]}(z)$ is a monic polynomial of degree
$n(n+1)/2$, then it can be written as
\begin{equation}
\label{6.4}Q_n^{[2]}(z)=\sum_{k=0}^{n(n+1)/2}A_{n,k}z^{n(n+1)/2-k},\qquad
A_{n,0}=1.
\end{equation}
Symmetric functions of its roots can be defined as it was done in
the case of the Yablonskii-Vorob'ev polynomials $Q_n(z)$ and the
polynomials $Q_n^{[2]}(z)$
\begin{equation}
\label{6.7}s_{n,m}\stackrel{def}{=}\sum_{k=1}^{n(n+1)/2}(a_{n,k})^m,\quad
m\geq 1,
\end{equation}
where $a_{n,k}$ $(1\leq k\leq n(n+1)/2)$ are the roots of
$Q_n^{[3]}(z)$. It can be proved the following theorems:

\begin{theorem}
\label{T:6.1} Let $c_{i,-m-1}$ be the coefficient in expansion
\eqref{6.3} at integer $\alpha=i\in \textbf{N}$. Then for each
$m\geq1$ and $n\geq2$ the following relation holds
\begin{equation}
\label{6.8}s_{n,m}=-\sum_{i=2}^{n}c_{i,-(m+1)}.
\end{equation}
\end{theorem}

\begin{theorem}
\label{T:6.2} All the coefficients $A_{n,m}$ of the polynomial
$Q_n^{[3]}(z)$ can be obtained with a help of $n(n+1)/2+1$ first
coefficients of the expansion \eqref{6.3} for the solutions of
\eqref{6.1}.
\end{theorem}

Thus we can calculate several functions $s_{n,m}$.
\begin{equation}\begin{gathered}
\label{kkk1}s_{{{n,7}}}=-\frac{5n}{2} \left( {n}^{2}-1 \right)
\left( {n}^{2}-4
 \right)  \left( {n}^{2}-9 \right)  \left( n+4 \right),
\end{gathered}\end{equation}

\begin{equation}\begin{gathered}
\label{kkk2}s_{{{n,14}}}=40\,n \left( {n}^{2}-1 \right) \left(
{n}^{2}-4
 \right)  \left( {n}^{2}-9 \right)  \left( n+4 \right) [5\,{n}^
{6}+15\,{n}^{5}-\\
\\
-340\,{n}^{4}-705\,{n}^{3}+8651\,{n}^{2}+9006\,n-77220],
\end{gathered}\end{equation}

\begin{equation}\begin{gathered}
\label{kkk3}s_{{{n,21}}}=-800\,n \left( {n}^{2}-1 \right) \left(
{n}^{2}-4
 \right)  \left( {n}^{2}-9 \right)  \left( n+4 \right)[35\,{n}
^{12}+210\,{n}^{11}-\\
\\
-6860\,{n}^{10}-36225\,{n}^{9}+629265\,{n}^{8}+
2736720\,{n}^{7}-32792630\,{n}^{6}-\\
\\
-108110865\,{n}^{5}+989372966\,{n}^{
4}+2162197152\,{n}^{3}-\\
\\
-16042160664\,{n}^{2}-17141744880\,n+ 107749699200],
\end{gathered}\end{equation}

\begin{equation}\begin{gathered}
\label{kkk4}s_{{{n,28}}}=144000\,n \left( {n}^{2}-1 \right) \left(
{n}^{2}-4
 \right)  \left( {n}^{2}-9 \right)  \left( n+4 \right) [ 35\,{n}
^{18}+315\,{n}^{17}-\\
\\
-13930\,{n}^{16}-118580\,{n}^{15}+2790550\,{n}^{14}
+21633990\,{n}^{13}-\\
\\
-351126160\,{n}^{12}-2393487040\,{n}^{11}+
29535328963\,{n}^{10}+\\
\\
+170146938815\,{n}^{9}-1682680983550\,{n}^{8}-
7778843727060\,{n}^{7}+\\
\\
+63914199838220\,{n}^{6}+219710453628328\,{n}^{5
}-1543528005776784\,{n}^{4}-\\
\\
-3462669359314848\,{n}^{3}+
21327037897395456\,{n}^{2}+\\
\\
+23096340571898880\,n-127579953840768000]
\end{gathered}\end{equation}

Now let us find the first few coefficients of $Q_n^{[3]}(z)$.
\begin{equation}\begin{gathered}
\label{kkk11}A_{{{n,7}}}={\frac {5}{14}}\,n \left( {n}^{2}-1 \right)
\left( {n} ^{2}-4 \right)  \left( {n}^{2}-9 \right)  \left( n+4
\right),
\end{gathered}\end{equation}

\begin{equation}\begin{gathered}
\label{kkk12}A_{{{n,14}}}={\frac {5}{392}}\,n \left( {n}^{2}-1
\right)  \left( { n}^{2}-4 \right)  \left( {n}^{2}-9 \right)  \left(
{n}^{2}-16 \right)
 \left( n+5 \right)\\
 \\
 [5\,{n}^{6}+15\,{n}^{5}-1105\,{n}^{4}-2235
\,{n}^{3}+56540\,{n}^{2}+57660\,n-864864],
\end{gathered}\end{equation}

\begin{equation}\begin{gathered}
\label{kk13}A_{{{n,21}}}={\frac {25}{16464}}\,n \left( {n}^{2}-1
\right)
 \left( {n}^{2}-4 \right)  \left( {n}^{2}-9 \right)  \left( {n}^{2}-16
 \right)  \left( {n}^{2}-25 \right) \\
 \\
 \left( n+6 \right) [5{n}
^{12}+30{n}^{11}-3235{n}^{10}-16450{n}^{9}+985395{n}^{8}+
4040610{n}^{7}-\\
\\
-137483057{n}^{6}-426660726{n}^{5}+9606229564{n}
^{4}+19928307448{n}^{3}-\\
\\
-331282163616\,{n}^{2}-341318105856\,n+ 4505374089216],
\end{gathered}\end{equation}

\begin{equation}\begin{gathered}
\label{kk14}A_{{{n,28}}}={\frac {25}{921984}}\,n \left( {n}^{2}-1
\right)
 \left( {n}^{2}-4 \right)  \left( {n}^{2}-9 \right)  \left( {n}^{2}-16
 \right)  \left( {n}^{2}-25 \right)\\
 \\
 \left( {n}^{2}-36 \right)(n+7)[25\,{n}^{18}+225\,{n}^{17}-31900\,{n}^{16}-
 260300\,{n}^{15}+\\
 \\
 +21092150\,{n}^{14}+152218150\,{n}^{13}-8908097220\,{n}^{12}-
55439329220\,{n}^{11}+\\
\\
+2295894377065\,{n}^{10}+11991324803825\,{n}^{9}-
362820266918120\,{n}^{8}-\\
\\
-1523845124481400\,{n}^{7}+35464390728403088\,
{n}^{6}+\\
\\
+111777610107420944\,{n}^{5}-2096257669340843520\,{n}^{4}-\\
\\
-4380613465733418240\,{n}^{3}+68742943019313426432\,{n}^{2}+\\
\\
+ 70952136972154454016\,n-960217764587156275200]
 \end{gathered}\end{equation}
Again rational solutions of \eqref{6.1} can be written in terms of
the logarithmic derivative of the polynomials $Q_n^{[3]}(z)$.

\section {Conclusion}

An alternative method for constructing the Yablonskii - Vorob'ev
polynomials has been presented. The basic idea of the method is to
use power expansions of solutions for the second Painlev\'{e}
equation. These power expansions can be found with a help of
algorithms of power geometry \cite{Bruno01, Bruno02}. Using our
approach we have derived formulas for the coefficients of the
Yablonskii - Vorob'ev polynomials and also we have obtained the
correlations between the roots of each polynomial. Our method can be
also applied for constructing polynomials associated with other
nonlinear differential equations \cite{Okamoto01, Clarkson04,
Kajiwara01, Nuomi01, Umemura01}.

\section {Acknowledgments}

Authors are thankful to professor V.V. Tsegel'nik for sending copies
of works \cite{Yablonskii01, Vorob'ev01}. This work was supported by
the International Science and Technology Center under Project B
1213.

\end{document}